\documentclass{article}
\usepackage{amssymb}

\input{tcilatex}

\begin{document}

\begin{center}
\textbf{UNRUH\ MODEL\ FOR\ THE\ EINSTEIN-ROSEN CHARGE: SQUEALING WORMHOLES?}

\bigskip

K.K. Nandi$^{a}$ and D.H. Xu$^{b}$

\bigskip \textit{Department of Physics, Zhejiang University of Technology,
Hangzhou 310032, China}
\end{center}

\bigskip

\begin{center}
\textbf{Abstract}
\end{center}

\ \ \ We present two kinds of acoustic models for the massless electric
charge conceived by Einstein and Rosen in the form of a bridge (wormhole
throat). It is found that the first kind of modelling requires a thin layer
of exotic matter at the bridge. We also derive an acoustic equation that
exclusively characterizes the model. Using a second kind of model, it is
demonstrated that the Einstein-Rosen charge has a sonic Hawking-Unruh
temperature proportional to $\mp 1/\beta $, where $\beta $ is the size of
the charge. This suggests that (squealing!) wormholes can also be formally
accommodated into Unruh's fluid model.

\bigskip

PACS number(s): 04.20.Gz, 04.60.+n, 47.90.+a, 97.60.Lf

------------------------------------------------------------------

E-mail address: $^{a}$kamalnandi@hotmail.com

E-mail address: $^{b}$xudh@zjut.edu.cn

\bigskip

\bigskip

\bigskip

\bigskip

\bigskip

\bigskip

\bigskip

\bigskip

\bigskip

\bigskip

\bigskip

\bigskip

\bigskip

\bigskip

\bigskip

\bigskip

\bigskip

\bigskip

\bigskip

\bigskip

\bigskip

\bigskip

\bigskip

\bigskip

\bigskip

\bigskip

\bigskip

\bigskip

\bigskip

\bigskip

\bigskip

\bigskip

\bigskip

\bigskip

\bigskip

\bigskip

\bigskip

In a recent paper, a first hand model of a class of static spherically
symmetric acoustic wormholes has been advanced as a natural completion of
the acoustic analog of black holes [1]. (The work contains the acoustic
analogs of Morris-Thorne wormholes with a minimally coupled scalar field
[2]). Acoustic models of black hole geometry have originated with Unruh's
novel discovery that a sonic horizon in transonic flow could give out a
thermal spectrum of sound waves mimicking Hawking's general relativistic
black hole evaporation [3]. The radiation of such sound waves is now
commonly known as Hawking-Unruh radiation [4]. This basic acoustic analogy
has been further explored under different physical circumstances by several
authors [5]. For a good discussion of acoustic geometries including the
Hawking-Unruh temperature, see Ref.[6]. The works surrounding the analogy
have engendered the practical possibility of detecting Hawking-Unruh
radiation in fluid (particularly in superfluid) models under appropriately
simulated conditions [7]. For this reason, the analog models have attracted
widespread attention as they open up alternative windows to look for many
unknown effects of quantum gravity on black holes in the laboratory.

However, the work reported here is somewhat limited in its scope but the
results could still be curious and useful. We shall begin by recalling a bit
of history. Einstein was not quite satisfied with the representation of
ponderable matter appearing in the general relativity field equations:
\textquotedblleft But it (general relativity) is similar to a building, one
wing of which is made of fine marble (the left side of the equation), but
the other wing of which is built of low grade wood (the right side of the
equation). The phenomenological representation of matter is, in fact, only a
crude substitute for a representation which would do justice to all known
properties of matter\textquotedblright\ [8]. In particular, Einstein was
averse to accepting material particles as singularities of the field. In an
effort to avoid these singularities, Einstein and Rosen [9] geometrically
modelled massive neutral elementary particles as \textquotedblleft
bridges\textquotedblright\ connecting two sheets of spacetime. These objects
are similar to what are now generally dubbed as Lorentzian wormholes and the
connecting bridges as wormhole throats. They also represented massless
charged particles as bridges in a similar fashion. The physicality of such
mathematical representations could always be arguable and our intention is
not to discuss it here. Instead, in the present Letter, we shall develop the
acoustic analogs of the massless electrical particle in two inequivalent
ways and show that sonic Hawking-Unruh temperature could be assigned to the
particle in a purely formal manner.

The construction of the acoustic analog is based on the identification of a
given general relativity metric with the acoustic metric derived from the
standard set of irrotational, nonrelativistic inviscid, barotropic fluid
equations given by

\begin{equation}
\overrightarrow{\nabla }\times \overrightarrow{v}=0\Rightarrow 
\overrightarrow{v}=\overrightarrow{\nabla }\Psi
\end{equation}

\begin{equation}
\frac{\partial \rho }{\partial t}+\overrightarrow{\nabla }.\left( \rho 
\overrightarrow{v}\right) =0
\end{equation}%
\begin{equation}
\rho \left[ \frac{\partial \overrightarrow{v}}{\partial t}+(\overrightarrow{v%
}.\overrightarrow{\nabla })\overrightarrow{v}\right] =-\overrightarrow{%
\nabla }p
\end{equation}

\begin{equation}
p=p(\rho )
\end{equation}%
in which all relevant terms have their usual meanings. Now linearize the
equations around a background exact solution set [ $p_{0}(t,\overrightarrow{x%
}),\rho _{0}(t,\overrightarrow{x}),\Psi _{0}(t,\overrightarrow{x})$] such
that%
\begin{equation}
p=p_{0}+\delta p_{0}+0(\delta p_{0})^{2},\rho =\rho _{0}+\delta \rho
_{0}+0(\delta \rho _{0})^{2},\Psi =\Psi _{0}+\delta \Psi _{0}+0(\delta \Psi
_{0})^{2}
\end{equation}%
where $\delta $ denotes a small perturbation to the relevant quantities.
Then the perturbations satisfy the equations [3]

\begin{equation}
-\partial _{t}\left[ \frac{\partial \rho }{\partial p}\rho _{0}\left(
\partial _{t}\Psi _{1}+\overrightarrow{v}_{0}.\overrightarrow{\nabla }\Psi
_{1}\right) \right] +\overrightarrow{\nabla }.\left[ \rho _{0}%
\overrightarrow{\nabla }\Psi _{1}-\frac{\partial \rho }{\partial p}\rho _{0}%
\overrightarrow{v}_{0}\left( \partial _{t}\Psi _{1}+\overrightarrow{v}_{0}.%
\overrightarrow{\nabla }\Psi _{1}\right) \right] =0.
\end{equation}%
This equation can be neatly rewritten as the minimally coupled wave equation
[6]

\begin{equation}
\frac{1}{\sqrt{-\overline{g}}}\frac{\partial }{\partial x^{\mu }}\left( 
\sqrt{-\overline{g}}\overline{g}^{\mu \nu }\frac{\partial \Psi _{1}}{%
\partial x^{\nu }}\right) =0
\end{equation}%
where $\overline{g}_{\mu \nu }$ is the acoustic metric, $\overrightarrow{v}%
_{0}\equiv \overrightarrow{\nabla }\Psi _{0}$, $\Psi _{1}\equiv \delta \Psi
_{0},$ $\overline{g}=\det \left\vert \overline{g}_{\mu \nu }\right\vert $,
and the stationary metric form is given by

\begin{equation}
ds_{acoustic}^{2}=\overline{g}_{\mu \nu }dx^{\mu }dx^{\nu }=\frac{\rho _{0}}{%
c_{s}}\left[ c_{s}^{2}dt^{2}-\left( dx^{i}-v_{0}^{i}dt\right) \delta
_{ij}\left( dx^{j}-v_{0}^{j}dt\right) \right]
\end{equation}%
in which $i,j=1,2,3$ and the local speed of sound is given by $c_{s}^{2}=%
\frac{\delta p_{0}}{\delta \rho _{0}}$. Before closing this part, we note
that in a physical situation where $\overrightarrow{v}_{0}=0$, and $\rho
_{0}/c_{s}$= constant, we have an exact Minkowski spacetime where the role
of signal speed is played by that of sound waves. On a scale larger than the
interatomic distance and in a reasonably local neighborhood inside the
fluid, this geometry underlies the basic ingredient of what one might call
sonic special relativity where the observers are equipped with only sound
waves, that is, a world where the observers can only \textquotedblleft
hear\textquotedblright\ but not \textquotedblleft see\textquotedblright . It
might be of some interest to note that a similar framework of sonic
relativity was conceived decades ago in which the subsonic and supersonic
Lorentz-like transformations containing $c_{s}$ as the invariant speed were
employed in the treatment of many acoustical problems in a very profitable
manner [10].

Proceeding further, note that the signature convention in use in this paper
is ($+,-,-,-$) and unless otherwise noted, we shall use units $8\pi G=c=1$.
Consider now the metric form used by Einstein and Rosen [9] in the standard
coordinates ($t,R,\theta ,\varphi )$:%
\begin{equation}
ds^{2}=\left( 1-\frac{2m}{R}-\frac{\epsilon ^{2}}{2R^{2}}\right)
dt^{2}-\left( 1-\frac{2m}{R}-\frac{\epsilon ^{2}}{2R^{2}}\right)
^{-1}dR^{2}-R^{2}d\theta ^{2}-R^{2}\sin ^{2}\theta d\varphi ^{2}
\end{equation}

\begin{equation}
\phi _{0}=\frac{\epsilon }{R}
\end{equation}%
where $\phi _{0}$ is the electrostatic potential while the magnetic
potentials $\phi _{i}$ are set to zero. An elementary electrical particle
without mass ($m=0$) is represented by the singularity free solution
obtained by the radial transformation ($R\rightarrow u)$ as follows:

\begin{equation}
u^{2}=R^{2}-\frac{\epsilon ^{2}}{2}.
\end{equation}%
The bridge (or the particle) is located at $u=0$ or $R_{b}=\epsilon /\sqrt{2}
$ connecting two sheets $u>0$ and $u<0$. (Similarly, a neutral particle
corresponds to $\epsilon =0$ with the radial transformation $u^{2}=R-2m$ so
that the particle is represented by a bridge at $u=0$.) For our purposes, we
shall rewrite the solution in isotropic coordinates ($t,r,\theta ,\varphi )$:

\begin{equation}
ds^{2}=g_{\mu \nu }dx^{\mu }dx^{\nu }=\Omega ^{2}(r)dt^{2}-\Phi ^{-2}(r) 
\left[ dr^{2}+r^{2}d\theta ^{2}+r^{2}\sin ^{2}\theta d\varphi ^{2}\right]
\end{equation}

\begin{equation}
\Omega ^{2}(r)=\left[ 1-\frac{m^{2}+\beta ^{2}}{4r^{2}}\right] ^{2}\left[ 1+%
\frac{m}{r}+\frac{m^{2}+\beta ^{2}}{4r^{2}}\right] ^{-2}
\end{equation}

\begin{equation}
\Phi ^{-2}(r)=\left[ 1+\frac{m}{r}+\frac{m^{2}+\beta ^{2}}{4r^{2}}\right]
^{2}
\end{equation}

\begin{equation}
\phi _{0}=\frac{4\sqrt{2}\beta r}{(m+2r)^{2}+\beta ^{2}}.
\end{equation}%
The massless bridge is now located at $r_{b}=\beta /2$. For the sake of
generality, we keep $m\neq 0$ for the moment. The factor $\sqrt{2\text{ }}$
is imported into Eq.(15) as we have slightly redefined the constant $%
\epsilon $ in Eq.(9) as $\epsilon ^{2}=2\beta ^{2}$ for the ease of
calculations. The arbitrary constants $m$ and $\beta $ represent,
respectively, the mass and electric charge of the configuration with the
solutions satisfying the field equations without denominators given by [9]

\begin{equation}
\phi _{\mu \nu }=\phi _{\mu ,\nu }-\phi _{\nu ,\mu }
\end{equation}

\begin{equation}
g^{2}\phi _{\mu \nu ;\sigma }g^{\nu \sigma }=0
\end{equation}

\begin{equation}
g^{2}(R_{\alpha \beta }+\phi _{\alpha \gamma }\phi _{\beta }^{\gamma }-\frac{%
1}{4}g_{\alpha \beta }\phi _{\lambda \delta }\phi ^{\lambda \delta })=0
\end{equation}%
in which $g=\det \left\vert g_{\mu \nu }\right\vert $, $\phi _{\mu }\equiv
(\phi _{0},\phi _{i})$ and $R_{\alpha \beta }$ is the Ricci tensor. The
metric (9) above with the term $-\epsilon ^{2}$ ($\epsilon ^{2}>0$) is not
quite the familiar Reissner-Nordstr\"{o}m metric which has $+\epsilon ^{2}$
in its place. Correspondingly, there appears an overall negative sign before
the stress energy tensor of the electrostatic field appearing on the right
hand side of the field equations. This, in turn, indicates that the
electrical stresses violate the energy conditions, as long as we hold $%
\epsilon ^{2}$ (or $\beta ^{2}$) $>0$ in the metric (9), thereby providing
materials necessary for the construction of the Einstein-Rosen bridge making
up charged particles. This violation (of energy conditions) is precisely the
price that one has to pay in order to build the bridge.

Let us return to the acoustic metric (8) and consider a preassigned velocity
profile $\overrightarrow{v}_{0}\neq 0$. If the vector $\overrightarrow{v}%
_{0}/(c_{s}^{2}-v_{0}^{2})$ is integrable, that is, $\overrightarrow{\nabla }%
\times \left[ \overrightarrow{v}_{0}/(c_{s}^{2}-v_{0}^{2})\right] =0$, then
defining a new time coordinate $d\tau =dt+[\overrightarrow{v}_{0}.d%
\overrightarrow{x}/(c_{s}^{2}-v_{0}^{2})]$, it is possible to write the
acoustic metric in the static form as [3b, 6]

\begin{equation}
ds_{acoustic}^{2}=\overline{g}_{\mu \nu }dx^{\mu }dx^{\nu }=\frac{\rho _{0}}{%
c_{s}}\left[ (c_{s}^{2}-v_{0}^{2})d\tau ^{2}-\left( \delta _{ij}+\frac{%
v_{0}^{i}v_{0}^{j}}{c_{s}^{2}-v_{0}^{2}}\right) dx^{i}dx^{j}\right] .
\end{equation}%
The first kind of acoustic model is obtained by choosing the simplest
velocity profile of the background fluid: $\overrightarrow{v}_{0}=0$. This
choice allows us to directly build the analogy in the fashion suggested by
Visser and Weinfurtner [11]: Replace the vacuum speed of light $c$ in (12)
[which we have set to unity] by the asymptotic speed of sound $c_{\infty }$
corresponding to a linear medium. Just a formal replacement; it does not
mean that numerically $c\equiv c_{\infty }$. Identify $t$ with $\tau $ and
the metric (12) with the metric (19), that is, set $g_{\mu \nu }\equiv $ $%
\overline{g}_{\mu \nu }.$ Immediately one obtains the density and sound
speed profiles respectively as

\begin{equation}
\rho _{0}(r)=\rho _{\infty }\left( 1-\frac{m^{2}+\beta ^{2}}{4r^{2}}\right)
\end{equation}

\begin{equation}
c_{s}(r)=c_{\infty }\left( 1-\frac{m^{2}+\beta ^{2}}{4r^{2}}\right) \left[ 1+%
\frac{m}{r}+\frac{m^{2}+\beta ^{2}}{4r^{2}}\right] ^{-2}
\end{equation}%
We are looking for an acoustic analog of the Einstein-Rosen charge which is
essentially a wormhole. Consequently, the analog fluid energy density $\rho
_{0}$ ($\geqslant 0$) is \textit{not} the actual energy density of wormhole
matter provided by the electrostatic field $\phi _{0}$, the latter density
being strictly negative (exotic) for $\beta ^{2}>0$, as already mentioned.
Returning to the fluid description, we can rewrite the Euler equation (3)
for $p_{0}=p_{0}(r)$ by explicitly displaying the force required to hold the
fluid configuration in place against the pressure gradient:

\begin{equation}
\overrightarrow{f}=\rho _{0}\left( \overrightarrow{v}_{0}.\overrightarrow{%
\nabla }\right) \overrightarrow{v}_{0}+c_{s}^{2}\partial _{r}\rho _{0}%
\widehat{r}.
\end{equation}%
It is now possible to find the pressure profile by integrating the equation:

\begin{equation}
f_{\widehat{r}}=\frac{dp_{0}}{dr}=c_{s}^{2}\frac{d\rho _{0}}{dr}
\end{equation}%
which gives, under the physical condition that $p_{\infty }\rightarrow 0$,

\[
p_{0}(r)=\frac{1}{12}\rho _{\infty }c_{\infty }^{2}\left( m^{2}+\beta
^{2}\right) [\frac{16(m^{4}+2m^{3}r-6mr\beta ^{2}-\beta ^{4})}{%
\{(m+2r)^{2}+\beta ^{2}\}^{3}} 
\]

\[
-\frac{4(m^{4}+2m^{3}r+3m^{2}\beta ^{2}-18mr\beta ^{2}-6\beta ^{4})}{\beta
^{2}\{(m+2r)^{2}+\beta ^{2}\}^{2}} 
\]

\[
-\frac{6(m^{4}+2m^{3}r+m^{2}\beta ^{2}+2mr\beta ^{2}+2\beta ^{4})}{\beta
^{4}\{(m+2r)^{2}+\beta ^{2}\}} 
\]

\begin{equation}
+\frac{3m(m^{2}+\beta ^{2})}{\beta ^{5}}\left( \pi -2\arctan \{\frac{m+2r}{%
\beta }\}\right) ].
\end{equation}%
The barotropic equation of state can be found by eliminating $r$ from
Eqs.(20) and (24), but we are not displaying it here. Instead we shall
concentrate on the $m=0$ case in accordance with the Einstein-Rosen model.
Then

\begin{equation}
p_{0}(r)=-\frac{\rho _{\infty }c_{\infty }^{2}\beta ^{2}(48r^{4}+\beta ^{4})%
}{3(4r^{2}+\beta ^{2})^{3}}.
\end{equation}%
Using the relation $r=\frac{\beta }{2\sqrt{1-\rho _{0}/\rho _{\infty }}}$,
we can write the equation of state as

\begin{equation}
p_{0}(\rho _{0})=-\rho _{\infty }c_{\infty }^{2}\frac{(1-\rho _{0}/\rho
_{\infty })(4-2\rho _{0}/\rho _{\infty }+\rho _{0}^{2}/\rho _{\infty }^{2})}{%
3(2-\rho _{0}/\rho _{\infty })^{3}}.
\end{equation}%
The radial force required to maintain the acoustic configuration is

\begin{equation}
f_{\widehat{r}}=\rho _{\infty }c_{\infty }^{2}\times \frac{\beta ^{2}}{2r^{3}%
}\times \frac{\left( 1-\frac{\beta ^{2}}{4r^{2}}\right) ^{2}}{\left( 1+\frac{%
\beta ^{2}}{4r^{2}}\right) ^{4}}
\end{equation}%
which is finite for all values of $r\neq 0$. It further turns out that $%
p_{0}(\rho _{0})$ blows up at $\rho _{0}/\rho _{\infty }=2$, but that is of
no concern as the radial variable $r$ is defined only for $\rho _{0}/\rho
_{\infty }<1$. At the bridge ($r_{b}=\beta /2$), we have $\rho
_{0}=0,c_{s}=0,$ and $p_{0}=-\rho _{\infty }c_{\infty }^{2}/6$. The
Eqs.(20), (21) with $m=0$, and Eqs.(26), (27) represent the exact acoustic
model for the Einstein-Rosen massless electrical particle that we have been
looking for. The following characteristics are observed. From the expression
(25), it follows that the pressure drops from zero at infinity to negative
values as one proceeds to the bridge radius. However, with regard to the
overall energy condition, normalizing $\rho _{\infty }=c_{\infty }=1$, we
can see that $\rho _{0}(r_{b})+p_{0}(r_{b})<0$ implying that the (dominant)
energy condition is violated at the bridge. Taking further $\beta =1$, that
is, $r_{b}=0.5$, we find graphically that at $r\approx 0.55,$ $\rho
_{0}+p_{0}=0$ and for $r>0.55$, we have $\rho _{0}+p_{0}>0$. Thus, there is
a thin layer of exotic acoustic material of the order of thickness $\delta
\approx 0.05$ wrapping up the bridge radius. Finally, we find that there is
a relation between the sound speed and density given by

\begin{equation}
\frac{c_{s}}{c_{\infty }}=\frac{\rho _{0}}{\rho _{\infty }}\left( 2-\frac{%
\rho _{0}}{\rho _{\infty }}\right) ^{-2}
\end{equation}%
which holds at every point in the medium characterizing the massless charge
in its acoustic model.

The second acoustic model may be framed as follows. Consider the $%
\overrightarrow{v}_{0}\neq 0$ situation together with the metric (19) and
note Unruh's original derivation [3]. The key assumptions were the constancy
of the sound speed $c_{s}$ (which we normalize to unity, but it can be set
to any other value as well) and a radial flow with a particular form of the
fluid velocity profile

\begin{equation}
v_{0}^{R}=-1+\alpha (R-R_{0})+O(R-R_{0})^{2}
\end{equation}%
in which $\alpha \lbrack \equiv (\partial v_{0}^{R}/\partial R)]$ evaluated
at $R=R_{0}$ is a constant proportional to sonic Hawking-Unruh temperature [$%
T_{H}=(\hslash \alpha /2\pi \kappa )$], $\kappa $ is Boltzmann constant and $%
R_{0}$ is the sonic horizon defined as the surface where $v_{0}^{R}=\pm 1$,
the minus sign denoting the opposite motions of the fluid and the sound.
Plugging Eq.(29) into the metric (19), and after dropping the angular parts,
it has the form

\begin{equation}
ds^{2}\approx \frac{\rho _{0}}{c_{s}}[2c_{s}\alpha (R-R_{0})d\tau ^{2}-\frac{%
dR^{2}}{2\alpha (R-R_{0})}]
\end{equation}%
which compares with the black hole metric near the horizon%
\begin{equation}
ds^{2}\approx \lbrack (\widehat{r}-2M)/2M]d\widehat{t}^{2}-2Md\widehat{r}%
^{2}/(\widehat{r}-2M)
\end{equation}%
provided one identifies $\alpha \equiv \frac{1}{4M}$.

Our acoustic configuration, on the other hand, has modelled a wormhole
spacetime given by (9), instead of a Schwarzschild black hole (for which $%
\beta =0$) and the two physical situations are entirely different [12]. For
instance, black holes are the possible end results of a collapse due to
Hawking-Penrose singularity theorems while these theorems do not apply when
energy conditions are violated. Nonetheless, using the assumption of
constancy of sound speed, and comparing the metrics (9) and (19), we get the
profiles $(v_{0}^{R})^{2}=2m/R+\beta ^{2}/R^{2}$, $\rho _{0}=1$, $p_{0}=0$.
This implies that we now have a dust-like fluid. The fluid horizon occurs at 
$R=R_{0}=m\pm \sqrt{m^{2}+\beta ^{2}}$ where $v_{0}^{R}=\pm 1$. Note from
the metric (9) that the wormhole is nontraversable due to the fact that the
throat radius $R_{th}$ coincides with the geometric horizon $R_{h}$ and that 
$R_{h}=R_{th}=R_{0}.$ By the Taylor Expansion of $v_{0}^{R}$ around $R=R_{0}$%
, as in Eq.(29), we find that $\alpha =\mp \frac{1}{R_{0}}(1-\frac{m}{R_{0}}%
) $. We can keep the sign open knowing that wormholes can have negative
Hawking temperatures as speculated by Hong and Kim [13] on different
grounds. The constants $m$, $\beta $ appearing in $\alpha $ and $R_{0}$ of
the fluid configuration must be treated as mere constants bereft of their
original physical meanings and it signifies the limitation of the present
approach. The main thing however is that $\alpha $ is \textit{finite} at the
throat and it can be related to the temperature of a massive electric charge
($m\neq 0,\beta \neq 0$). Within the framework embodied in Eqs.(30) and
(31), an equivalent\textit{\ }description can be assigned to the wormhole in
terms of a fictitious Schwarzschild horizon for mass $M$ given by $M=\pm 
\frac{R_{0}^{2}}{4(R_{0}-m)}$. In the Einstein-Rosen model of neutral
particle ($\beta =0$), the value of sonic $T_{H}$ is proportional to $\mp
1/4m$, as expected, while for the massless ($m=0$) electrical particle, it
is proportional to $\mp 1/\beta $. All these results have followed from a
direct application of Unruh's method with the only physical difference that
the metric (9) represents a wormhole, \textit{not} a black hole. If, on the
other hand, we had $\beta ^{2}<0$ (that is, the usual Reissner-Nordstr\"{o}m
case with positive stresses), then $R_{0}$ or the sonic $T_{H}$ would have
been imaginary for $m=0$. This indicates that the existence of real sonic $%
T_{H}$ \ is consistent more with exotic matter present in spacetime than
with ordinary matter at least in the case under discussion here.

Finally, one might write down the acoustic equations corresponding to the
general Morris-Thorne wormhole given by

\begin{equation}
ds^{2}=e^{2\Phi (R)}dt^{2}-\frac{dR^{2}}{1-b(R)/R}-R^{2}d\theta
^{2}-R^{2}\sin ^{2}\theta d\varphi ^{2}
\end{equation}%
where $\Phi (R)$ and $b(R)$ are the redshift and shape functions,
respectively, satisfying the appropriate conditions. The throat is defined
at $R=R_{0}$ by $b(R_{0})=R_{0}$. A procedure similar to that in the above
paragraph leads to the exact expressions for density, radial velocity and
its gradient as follows:

\begin{equation}
\rho _{0}(R)=e^{\Phi }\left( 1-\frac{b}{R}\right) ^{-\frac{1}{2}}
\end{equation}

\begin{equation}
(v_{0}^{R})^{2}=1-e^{\Phi }\left( 1-\frac{b}{R}\right) ^{\frac{1}{2}}
\end{equation}

\begin{equation}
\frac{dv_{0}^{R}}{dR}=-\frac{1}{2v_{0}^{R}}\left[ \left( 1-\frac{b}{R}%
\right) ^{\frac{1}{2}}\frac{d(e^{\Phi })}{dR}+e^{\Phi }\frac{d}{dR}\left( 1-%
\frac{b}{R}\right) ^{\frac{1}{2}}\right] .
\end{equation}%
This shows that at $R=R_{0}$, we have a fluid horizon defined by $%
v_{0}^{R}=\pm 1$, but the gradient $\alpha \equiv \frac{dv_{0}^{R}}{dR}%
\rightarrow \infty $ \ due to the last term thereby rendering the notion of
sonic Hawking radiation meaningless. It might be possible to choose $\Phi $
and $b$ in such a way as to offset this divergence like the Einstein-Rosen
case.

\ \ The results in this paper have been derived in quite straightforward way
using a \textquotedblleft back door\textquotedblright\ approach that has its
own limitations. Nonetheless, the analyses above have attempted to provide
some curious insights as to how acoustic analogs of the Einstein-Rosen
charge (actually a wormhole) would look like. The conclusions are
essentially of academic interest at present knowing that the corresponding
sonic configurations may even be physically absurd. The important lesson
seems to be that the Hawking-Unruh radiation (involving the existence of
sonic horizon) need not emanate from a dumbhole alone; it can, in principle,
emanate also from the throat of a wormhole threaded by nontrivial exotic
matter \textit{reinterpreted} as a fictitious Schwarzschild mass $M$. That
is, worms can squeal - acoustically, that is! However, a few words of
caution are necessary. As shown recently, the possibility of actual black
hole evaporation itself depends on several crucial assumptions if it has to
be independent of the laws of Planck scale physics [14]. The situation in
the context of wormholes seems to be worse. The Ford-Roman quantum
inequality [15] suggests that the wormhole sizes could be tiny but the
mechanism of any wormhole evaporation is altogether unknown as these objects
are not formed in a collapse process. All these circumstances considerably
complicate, if not nullify, the very foundation of the acoustic
reincarnations of Hawking radiation from wormholes. Thus the ultimate
validity of acoustic analogies playing the role of alternative black or
wormholes must come from the successes of appropriately designed
acoustics/optics experiments in the laboratory.\ At any rate, we have
provided two inequivalent acoustic models of Einstein-Rosen massless charge
and found that, in the first case ($\overrightarrow{v}_{0}=0)$, one needs a
thin layer of exotic fluid for its construction. This exotic behavior
together with the Eq.(28) seem to characterize the model in an interesting
way. The second analog of the charge ($\overrightarrow{v}_{0}\neq 0$) shows
up a sonic Hawking temperature proportional to $\mp 1/\beta .$

\begin{center}
\textbf{Acknowlegments}
\end{center}

One of us (KKN) is indebted to Professor Liu Liao of Beijing Normal
University, China for encouragement in the present work. He thanks the
authorities of the University of North Bengal, India for granting leave.

\begin{center}
\textbf{References}
\end{center}

\bigskip \lbrack 1] Kamal Kanti Nandi, Yuan-Zhong Zhang, and Rong-Gen Cai,
[arXiv:gr-qc/0409085].

[2] K.K. Nandi and Yuan-Zhong Zhang, Phys. Rev. D \textbf{70}, 044040
(2004); K.K. Nandi, Yuan-Zhong Zhang, and K.B. Vijaya Kumar, Phys. Rev. D 
\textbf{70}, 064018 (2004); K.K. Nandi, A. Islam and J. Evans, Phys. Rev. D 
\textbf{55}, 2497 (1997). K.K. Nandi, B. Bhattacharjee, S.M.K. Alam and J.
Evans, Phys. Rev. D \textbf{57}, 823 (1998).

[3] W.G. Unruh, Phys. Rev. Lett. \textbf{46}, 1351 (1981); Phys. Rev. D 
\textbf{51}, 2827 (1995).

[4] W.G. Unruh, Phys. Rev. D \textbf{14}, 870 (1976). See, for the latest
reference, and a simpler derivation: Paul M. Alsing and Peter W. Milonni,
Am. J. \ Phys. (to appear in Nov. 2004), [arXiv:quant-ph/0401170].

[5] The literature is too vast to be exhaustive. The articles in the
following recent book convey the present state of affairs: M. Novello, M.
Visser and G. Volovik, (Eds): \textit{Artificial Black Holes}, (World
Scientific, Singapore, 2002); For a critique, a new process of regeneration
and some open questions regarding black hole Hawking radiation, see: T.
Jacobson, Phys. Rev. D \textbf{44}, 1731 (1991). For several other latter
works, see: T. Jacobson, Phys. Rev. D \textbf{48}, 728 (1993); \textit{ibid}%
. D \textbf{53}, 7082 (1996); T. Jacobson and G.E. Volovik, JETP Lett. 
\textbf{68}, 874 (1998); C. Barcel\'{o}, S. Liberati, S. Sonego and M.
Visser, [arXiv:gr-qc/0408022]; C. Barcel\'{o}, S. Liberati, and M. Visser,
Int. J. Mod. Phys. D \textbf{12}, \ 1641 (2003); Phys. Rev. A \textbf{68},
053613 (2003); M. Visser, Phys. Rev. D \textbf{48}, 583 (1993);\textit{\ ibid%
}. D \textbf{48}, 5697 (1993); Phys. Rev. Lett. \textbf{80}, 3436 (1998); 
\textit{ibid}. \textbf{85}, 5252 (2000); S. Corley, Phys. Rev. D \textbf{55}%
, 6155 (1997); S. Corley and T. Jacobson, Phys. Rev. D \textbf{54}, 1568
(1996); \textit{ibid}. D \textbf{57}, 6269 (1997); G.E. Volovik and T.
Vachaspati, Int. J. Mod. Phys. B \textbf{10}, 471 (1996); G.E. Volovik,
[arXiv:cond-mat/9706172]; B. Reznik, Phys. Rev. D \textbf{55}, 2152 (1997);
B. Reznik, [arXiv:gr-qc/9703076].

[6] M. Visser, Class. Quant. Grav. \textbf{15}, 1767 (1998).

[7] G.E. Volovik, \textit{Universe in a Helium Droplet} (Oxford University
Press, Oxford, 2003); L.J. Garay, J.R. Anglin, J.I. Cirac, and P. Zoller,
Phys. Rev. Lett. \textbf{85}, 4643 (2003); Phys. Rev. A \textbf{63}, 023611
(2001); C. Barcel\'{o}, S. Liberati, and M. Visser, Class. Quant. Grav. 
\textbf{18}, 1137 (2001); M.V. Berry, R.G. Chambers, M.D. Large, C. Upstill,
and J. Walmsley, Euro. J. Phys. \textbf{1}, 154 (1980); P. Roux, J. de
Rosny, M. Tanter, and M. Fink, Phys. Rev. Lett. \textbf{79}, 3170 (1997);
For optical black holes and slow light, see: L.V. Hau, S.E. Harris, Z.
Dutton, and C.H. Behroozi, Nature (London) \textbf{397}, 594 (1999); H.
Davidowitz and V. Steinberg, Europhys. Lett. \textbf{38}, 297 (1997); U.
Leonhardt and P. Piwnicki, Phys. Rev. Lett. \textbf{84}, 822 (2000); \textit{%
ibid}, \textbf{85}, 5253 (2000); U. Leonhardt, Phys. Rev. A \textbf{65},
043818 (2002) and references therein; I. Brevik and G. Halnes, Phys. Rev. D 
\textbf{65}, 024005 (2002); R. Sch\"{u}tzhold, G. Plunien, and G. Soff,
Phys. Rev. Lett. \textbf{88}, 061101 (2002).

[8] Recently quoted in T.A. Roman, [arXiv:gr-qc/0409090]: A. Einstein from 
\textit{Physics and Reality} (1936), reprinted in \textit{Ideas and Opinions}
(Crown, New York, 1954), p.311.

[9] A. Einstein and N. Rosen, Phys. Rev. \textbf{48}, 73 (1935).

[10] T.S. Shankara and Kamal Kanti Nandi, J. Appl. Phys. \textbf{49}, 5783
(1978); J. Tech. Phys. Polish Acad. Sci. \textbf{27}, 399 (1986).

[11] M. Visser and S.E.Ch. Weinfurtner, [arXiv:gr-qc/0409014].

[12] The Schwarzschild black hole can also be interpreted as a
nontraversable Morris-Thorne wormhole for which the throat and horizon radii
coincide. However, the interpretation is more of a technical nature as the
spacetime is empty unlike the case $\beta ^{2}>0$ for which there \textit{is}
nontrivial exotic material provided by the electrical stresses.

[13] S-T. Hong and S-W. Kim, [arXiv:gr-qc/0303059]. The negative temperature
is referred to also by T.A. Roman in Ref.[8] and P.F. Gonz\'{a}lez-D\'{\i}az
and C.L. Sig\"{u}enza, Nucl. Phys. B \textbf{697}, 363 (2004)
[arXiv:astro-ph/0407421]; P.F. Gonz\'{a}lez-D\'{\i}az and C.L. Sig\"{u}enza,
Phys. Lett. B \textbf{589}, 78 (2004).

[14] A very recent article argues that the phenomenon of black hole Hawking
radiation may still be an open question: W.G. Unruh and Ralf Sch\"{u}tzhold,
[arXix:gr-qc/0408009].

[15] L.H. Ford and T.A. Roman, Phys. Rev. D \textbf{53}, 5496 (1996). A
comprehensive recent discussion can be found in T.A. Roman, Ref. [8] and
[2b].

\bigskip

\end{document}